\documentclass[prb,reprint,aps,amsmath,amssymb,showpacs]{revtex4-1}

\usepackage{graphicx}

\begin{document}
\preprint{}
\draft

\title{Spin-based single-photon transistor, dynamic random access memory, diodes and routers in semiconductors}

\author{C.Y.~Hu}\email{chengyong.hu@bristol.ac.uk}
\affiliation{Department of Electrical and Electronic Engineering,
Merchant Venturers School of Engineering, Faculty of Engineering, University of Bristol,
Woodland Road, Bristol, BS8 1UB, United Kingdom}

\begin{abstract}

The realization of quantum computers and quantum Internet requires not only quantum
gates and quantum memories, but also transistors at single-photon levels to control
the flow of information encoded on single photons. Single-photon transistor (SPT) is
an optical transistor in the quantum limit, which uses a single photon to open or
block a photonic channel. In sharp contrast to all previous SPT proposals which are
based on single-photon nonlinearities, here I present a novel design for a high-gain and
high-speed (up to THz) SPT based on a linear optical effect -  giant circular birefringence (GCB)
induced by a single spin in a double-sided optical microcavity. A gate photon sets
the spin state via projective measurement and controls the light propagation
in the optical channel. This spin-cavity transistor can be directly configured as
diodes, routers, DRAM units, switches, modulators, etc. Due to the duality as quantum
gate and transistor, the spin-cavity unit provides a solid-state platform ideal
for future Internet - a mixture of all-optical Internet with quantum Internet.

\end{abstract}

\date{\today}


\maketitle

\section{Introduction}

The invention of Internet has changed our daily lives so widely and deeply, and this trend
is accelerating with the recent progress in big data and cloud computing.
Although the current Internet is already very fast and flexible, it is neither very secure nor
very energy efficient.
The regular Internet uses light pulses to transmit information across
fiber-optic networks. These classical optical pulses can be easily intercepted and copied by
a third party without any alert.  Quantum Internet \cite{kimble08} with unconditional security
uses individual quanta of light - photons - to encode and transmit information. Photons can
not be measured without being destroyed due to the laws of quantum mechanics, \cite{wootters82, bouwmeester00, nielsen00}
so any kind of hacking can be monitored and evaded. The future Internet is very likely the
mixture of regular Internet and quantum Internet. The regular Internet would be used by
default, but switched over to quantum Internet when sensitive data need to be transmitted.
Moreover, the current Internet is not fully transparent and continues to employ electronic
information processing and energy-consuming optical-electrical /electrical-optical conversions \cite{saleh12}
as the long-sought optical information processing \cite{miller10} and optical buffering \cite{tucker05, burmeister08}
are not available yet, which hinders the development of all-optical networks. \cite{saleh12}

As an optical transistor in the quantum limit, SPT is a remarkable device that could build
a bridge between quantum networks and all-optical networks. Several SPT prototypes \cite{chang07, faraon08, volz12, reinhard12, hwang09, chen13, gorniaczyk14, tiarks14}
have been proposed recently, all exploiting single-photon nonlinearities, i.e., photon-photon interactions. \cite{chang14}
However, photons do not interact with each other intrinsically, so indirect photon-photon interactions via
electromagnetically induced transparency (EIT), \cite{fleischhauer05} photon blockade \cite{imamoglu97, tian92}
and Rydberg blockade \cite{lukin01} have been investigated in this context since last two decades in both natural atoms \cite{birnbaum05, dayan08, schuster08, kubanek08}
and artificial atoms including superconducting boxes \cite{fink08, deppe08, bishop09, lang11} and quantum dots (QDs). \cite{srinivasan07, faraon08, kasprzak10, volz12, reinhard12}
The QD cavity QED is a promising solid-state platform for information and communication
technology (ICT) due to their inherent scalability and mature semiconductor technology.
However, the photon blockade resulting from the anharmonicity of Jaynes-Cummings energy ladder \cite{jaynes63}
is hard to achieve due to the small ratio of the QD-cavity coupling strength to the
system dissipation rates \cite{reithmaier04, yoshie04, peter05, hennessy07, srinivasan07,  faraon08, kasprzak10, ohta11, volz12}
compared with other systems. \cite{birnbaum05, dayan08, schuster08, kubanek08, fink08, deppe08, bishop09, lang11}
Moreover, the gain of this SPT based on photon blockade is quite limited and only $2.2$ is expected
for In(Ga)As QDs. \cite{faraon08, volz12}

In this work I propose a different SPT scheme, which exploits photon-spin interactions rather
than photon-photon interactions in a QD-cavity system. This spin-cavity transistor is a genuine quantum transistor
in three aspects: (1) it is based on a quantum effect, i.e., the linear GFR; (2) it has  the duality as a quantum gate for QIP
and a classical transistor for OIP; (3) it can work in the quantum limit as a SPT to amplify a single-photon state to
Schr\"{o}dinger cat state. Therefore this new transistor can be more powerful than the traditional electronic transistors.
Theoretically the maximum gain
can reach $\sim10^5$ in the state-of-the-art pillar microcavity depending on the QD-cavity
coupling strength and the spin coherence time. The large gain is attributed to the linear GCB
that is robust against classical and quantum fluctuations. The speed which is determined
by the cavity lifetime has the potential to break the THz barrier for electronic
transistors. \cite{schwierz07, deal11} Thanks to the linear GCB, this SPT is genuinely a quantum
transistor with the duality as a quantum gate for quantum information processing and a transistor
for classical information processing, thus it could be more powerful than the conventional transistors.
Based on this versatile spin-cavity unit, quantum computers, \cite{nielsen00, ladd10} quantum
Internet \cite{kimble08} and high-speed (up to THz)  optical information processing \cite{miller10}
and optical buffering \cite{tucker05, burmeister08} can be realized with current semiconductor technology.

This work is organized as follows: In Sec. II, I give a brief discussion on the linear GCB induced
by a single QD-confined spin in a double-sided optical microcavity and its application for robust
quantum gate against various quantum and classical fluctuations. In Sec. III, I demonstrate that the linear GCB
offers a new amplification mechanism for a SPT. Applications of this transistor are discussed in Sec. IV for
DRAM units, Sec. V for diodes and isolators, and Sec. VI for routers.
Conclusions and outlook are presented in Sec. VII.

\section{Linear GCB for robust quantum gate}
A single electron or hole spin confined in a charged QD in an optical microcavity can induce
macroscopic GCB.\cite{hu09} The linear GCB \cite{hu15} is a linear optical effect that is responsible for
robust quantum gate operation discussed in this Section and SPT operation in next Section.
Fig. 1(a) shows such a spin-cavity unit with a negatively
charged QD in a double-sided symmetric pillar microcavity.
This type of cavity  \cite{gerard96, schneider16} can be fabricated from a planar
microcavity defined by two distributed Bragg reflectors (DBRs) with the cavity length chosen to
be one wavelength $\lambda$ such that the cavity field maxima lies at the center of the microcavity.
Both DBRs are partially reflective allowing the external light to couple in and out of the cavity,
and are symmetric to achieve unity resonant transmission when the charged QD decouples to the cavity.
Three-dimensional confinement of light in a pillar microcavity is provided by the two DBRs and
additional transverse index guiding. The cross section of the pillar cavity is made circular in order
to support circularly polarized light. Some photonic crystal nanocavities with specific symmetry
could support circularly polarized light and are suitable for this work as well. \cite{vahala03}
However, the advantage to use pillar microcavity is the high coupling efficiency as the fundamental
cavity mode is gaussian-like and matches perfectly with the external laser beam. The cavity
mode is designed to be in resonance with the optical transitions of QDs.

\begin{figure}[ht]
\centering
\includegraphics* [bb= 112 160 527 557, clip, width=7cm, height=7cm]{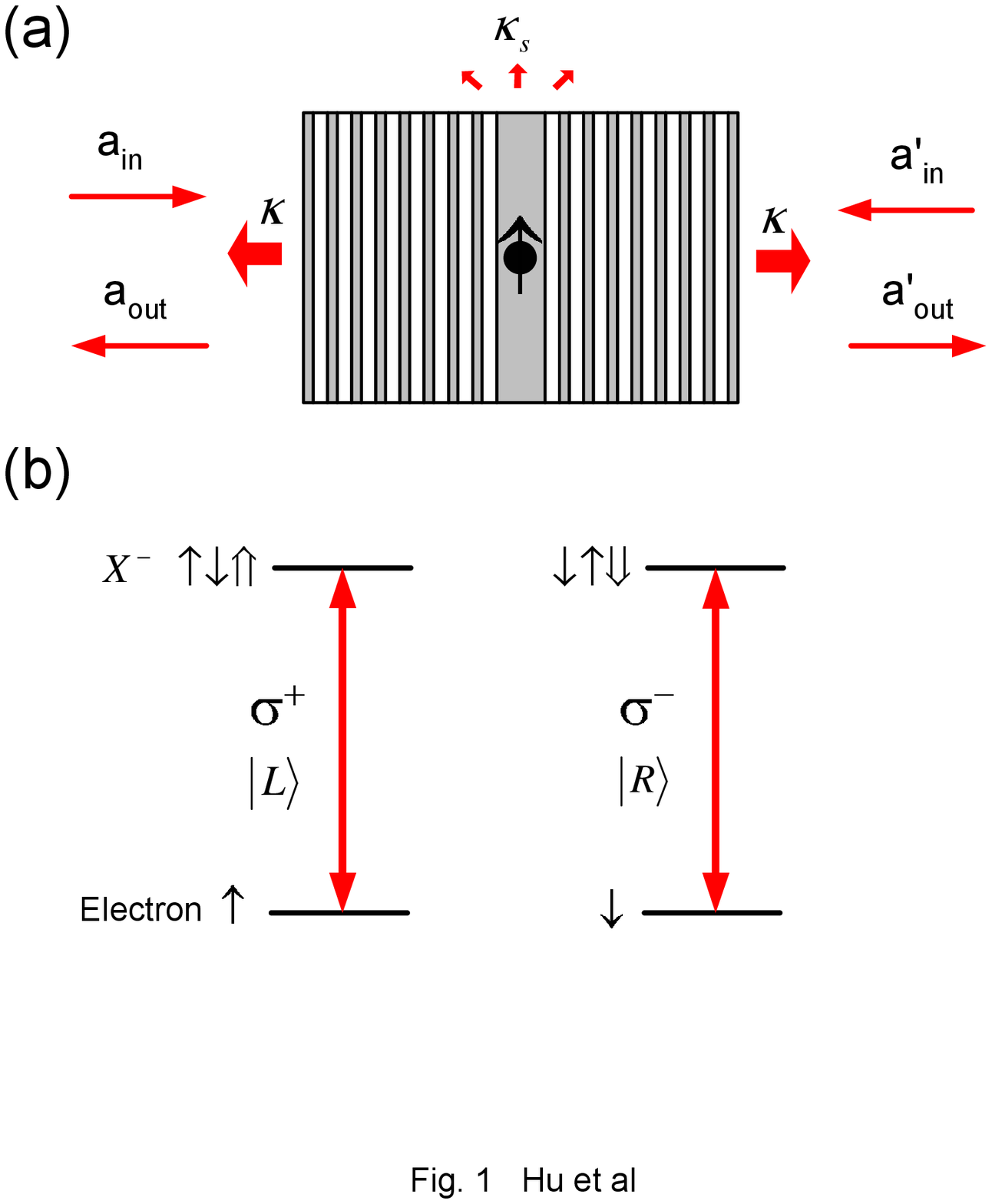}
\caption{(color online). Structure of the spin-cavity unit.  (a) A charged
quantum dot is embedded in a double-sided pillar microcavity with two symmetric
distributed Bragg reflectors allowing unity resonant transmission
when the charged quantum dot decouples to the cavity. The circular pillar cross-section
supports circularly polarized light.
(b) Optical transitions in a negatively-charged quantum dot follow spin selection rules: a photon
in the $|L\rangle$ state couples to the transition $|\uparrow\rangle \leftrightarrow |\uparrow\downarrow\Uparrow\rangle$ only,
whereas a photon in the $|R\rangle$ state couples to
the transition $|\downarrow\rangle \leftrightarrow |\downarrow\uparrow\Downarrow\rangle$ only due to
the conservation of angular momentum and the Pauli exclusion principle.}
\label{fig1}
\end{figure}

A negatively (or positively) charged QD has an excess electron (or hole ) confined in the QD. Charging a QD
can be achieved by modulation doping techniques, or tunneling in the n-i-n structures. \cite{warburton13}
The ground states of the charged QD are the electron (or hole) spin states, and the excited states are the
spin states of the negatively charged exciton $X^-$ [see Fig. 1(b)]. In the absence of external magnetic
field, both the ground and excited states of charged QD are two-fold degenerate due to the Kramers theorem.
The electron spin degeneracy could be lifted by the nuclear spin magnetic fields \cite{hansom14, urbaszek13}
via the electron-nucleus hyperfine interactions in In(Ga)As QDs, however, the Zeeman splitting is too small
to spoil the linear GCB. \cite{hu15} The hole spin degeneracy is not affected by the nuclear spin fields
due to the lack of the hole-nucleus hyperfine interactions.

Due to the conservation of total spin angular momentum and the Pauli exclusion principle, the left circularly
polarized photon (marked by  $|L\rangle$ or $|\sigma^+\rangle$) only couples to the transition
$|\uparrow\rangle \leftrightarrow |\uparrow\downarrow\Uparrow\rangle$, and the right circularly polarized
photon (marked by  $|R\rangle$ or $|\sigma^-\rangle$) only couples to the transition
$|\downarrow\rangle \leftrightarrow |\downarrow\uparrow\Downarrow\rangle$ [see Fig. 1(b)].
Here $|\uparrow\rangle$  and $|\downarrow\rangle$ represent electron spin states
$|\pm \frac{1}{2}\rangle$, $|\Uparrow\rangle$ and $|\Downarrow\rangle$ represent heavy-hole spin
states $|\pm\frac{3}{2}\rangle$ with the spin quantization axis z along QD growth direction, i.e.,
the input/output direction of light. The weak cross transitions due to the heavy-hole-light-hole mixing
can be corrected \cite{liu10} and are neglected in this work. Note that the photon polarizations are marked by
the input states to avoid any confusion due to the temporary polarization changes upon reflection.

If the spin is in the $|\uparrow\rangle$ state,
a photon in the $|L\rangle$ state can couple to the QD  due to the conservation of total spin
angular momentum and feels  a \textquotedblleft hot\textquotedblright cavity, whereas a photon in
the $|R\rangle$ state can not couple to the QD due to the Pauli exclusion principle and feels
a \textquotedblleft cold\textquotedblright cavity [see Fig. 1(b)]. If the spin is in the
$|\downarrow\rangle$ state, a $|R\rangle$-photon feels a \textquotedblleft hot\textquotedblright
cavity and a  $|L\rangle$-photon feels a \textquotedblleft cold\textquotedblright cavity.

The reflection/transmission coefficients of the hot and cold cavity are different as the QD-cavity
interactions could modify the cavity properties. This cavity-QED effect is verified by the calculations
with two approaches: an analytical method by solving Heisenberg-Langevin equations of
motions in the semi-classical approximation (see Appendix A), and a numerical but exact method by
solving master equation (see Appendix B) with a quantum optics toolbox. \cite{tan99, johansson13}
The calculated results were presented and discussed in Ref. \onlinecite{hu15}.

The different reflection/transmission coefficients between the hot and cold cavity lead to GCB
between two circular polarizations.\cite{hu09} GCB can be regarded as a macroscopic imprint of the microscopic
spin selection rules of charged QD as shown in Fig. 1(b). GCB is a type of magnetic optical
gyrotropy (also known as magnetic optical activity) in the presence of magnetic field or magnetization. \cite{landau60}
A key feature of GCB is its spin tunability, which makes the spin-cavity unit versatile for various
quantum or classical optical devices as demonstrated in this work. Another merit is the linear GCB
that remains constant with increasing the input-light power. \cite{hu15}

The linear GCB occurs around the cavity resonance in the strong coupling regime $\mathrm{g} \gg (2\kappa+\kappa_s, \gamma)$
or in the Purcell regime  $\gamma < 4\mathrm{g}^2/(2\kappa+\kappa_s)< (2\kappa+\kappa_s)$
when the input power is less than $P_{max}$ such that the QD stays in the ground state. In this case,
the semi-classical approximation can be used the coherent scattering dominate the reflection /transmission
processes (see Appendix A and Ref. \onlinecite{hu15}). Taking $\langle\sigma_z\rangle=-1$, the steady-state
reflection and transmission coefficients of the cavity can be obtained from Eq. (A2) in Appendix A.
\begin{equation}
\begin{split}
& r(\omega)=1+t(\omega), \\
& t(\omega)=\frac{-\kappa[i(\omega_{X^-}-\omega)+\frac{\gamma}{2}]}{[i(\omega_{X^-}-\omega)+
\frac{\gamma}{2}][i(\omega_c-\omega)+\kappa+\frac{\kappa_s}{2}]+\text{g}^2},
\end{split}
\label{eqd0}
\end{equation}
where $\omega$, $\omega_c$, $\omega_{X^-}$ are the frequencies of input light, cavity mode, and the $X^-$ transition,
respectively.  g is the QD-cavity coupling strength. $\kappa/2$ is the the cavity field decay rate into the input/output
port, and  $\kappa_s/2$ is the cavity field side leakage rate with the material background absorption included.
 $\gamma/2$ is the total QD dipole decay rate including the spontaneous emission rate $\gamma_{\parallel}/2$
into leaky modes and the pure dephasing rate $\gamma^*$, i.e., $\gamma/2=\gamma_{\parallel}/2+\gamma^*$.
The pure dephasing rate can be neglected when the QD is the ground state, which was proved in recent experiments
on high-quality single photon emission in In(Ga)As QDs under weak resonant excitation. \cite{nguyen11, matthiesen12, ding16, somaschi16, unsleber16}
For pillar microcavity, the spontaneous emission rate into leaky modes is approximately equal to the free-space
emission rate \cite{gerard98, solomon01} as the reduced density state of leaky modes can be compensated by the
Purcell enhancement. \cite{bjork94}

The resonant condition $\omega_c=\omega_{X^-}=\omega_0$ is considered in this work.
In the one-dimensional atom regime where $4\mathrm{g}^2 \gg (2\kappa+\kappa_s)\gamma$ (this includes
the strong coupling regime and part of the Purcell regime), equation (\ref{eqd0})
yields $r_h(\omega_0)\simeq 1$ and  $t_h(\omega_0)\simeq 0$ for the hot cavity.
If the side leakage is smaller than the input/output coupling rate, i.e.,  $\kappa_s\ll \kappa$,
$t_0(\omega_0)\simeq -1$ and $r_0(\omega_0)\simeq 0$ for the cold cavity. This property together
with the spin tunability enables a deterministic photon-spin entangling gate (or entanglement
beam splitter) with the transmission and reflection operators defined as \cite{hu09}
\begin{equation}
\begin{split}
&\hat{t}(\omega_0)=-(|R\rangle\langle R|\otimes |\uparrow \rangle\langle \uparrow|+|L\rangle\langle L|\otimes
|\downarrow \rangle\langle \downarrow|)\\
&\hat{r}(\omega_0)=|R\rangle\langle R|\otimes |\downarrow \rangle\langle \downarrow|+|L\rangle\langle L|\otimes
|\uparrow \rangle\langle \uparrow|.
\end{split} \label{ebs}
\end{equation}
This quantum gate can directly split a photon-spin polarization product state into two constituent
photon-spin entangled states with high fidelity in the strong coupling regime and the Purcell regime.
The larger is the QD-cavity coupling strength, the higher is the fidelity. \cite{hu09} Recently, strongly coupled
QD-cavity systems have been demonstrated in various micro- or
nano-cavities. \cite{reithmaier04, yoshie04, peter05, hennessy07, srinivasan07, faraon08, kasprzak10, volz12}
In the state-of-the-art pillar microcavities, \cite{reithmaier04, reitzenstein07}
$\mathrm{g}/(2\kappa+\kappa_s)=2.4$ is achievable for In(Ga)As QDs and is used for judging
the device performance in this work. Significant progress has been achieved towards the practical
implementation of the proposed photon-spin entangling gate,
e.g., the demonstration of a photon sorter, \cite{bennett16} a quantum switch, \cite{sun16}
Faraday rotation of $6^{\circ}$ induced by a single hole spin  \cite{arnold15} or electron spin. \cite{androvi16}
However, these experiments were performed in weakly-coupled cavity-QED systems \cite{waks06, an09, bonato10} (similar
to waveguide-QED structures \cite{shen05, lodahl15}) with lower device
performance, e.g., suffering from spectral diffusion of QD, low $P_{max}$, low or no SPT gain,
and vulnerable to quantum or classical fluctuations and electric /magnetic fields. Strongly-coupled cavity-QED
systems \cite{reithmaier04, yoshie04, peter05, hennessy07, srinivasan07, faraon08, kasprzak10, volz12}
would allow quantum gate \cite{duan04, hu09, hu15} and SPT with high performance.

For the linear GCB, the QD stay in the ground state, such that the photon-spin quantum gate is robust against quantum
fluctuations, \cite{hu15} such as the intensity fluctuations of incoming light. If working in the strong
coupling regime, this gate is also resistant to external/internal electrical /magnetic fields,
spectral diffusion, \cite{konthasinghe12} pure dephasing, \cite{greuter15} nuclear spin fluctuations \cite{hansom14, urbaszek13}
and even non-Markovian processes, \cite{kaer10} all of which could occur in realistic QDs. The non-saturation
window in the strong coupling regime protects the linear GCB from both classical and quantum fluctuations, \cite{hu15}
and leads to robust quantum gate and transistor operations.

The photon-spin entangling gate itself can be used to initialize the spin via single-photon based spin
projective measurement together with classical optical pulses injected from the cavity side. Assume the
spin is in a unknown state $\alpha |\uparrow \rangle +\beta |\downarrow \rangle$, and the photon in
the $|R\rangle$ state. After the photon-spin interaction, the photon and spin become entangled, i.e.,
$\alpha |R\rangle^t|\uparrow \rangle+\beta |R\rangle^r|\downarrow \rangle$. On detecting a transmitted
photon, the spin is projected to  $|\uparrow \rangle$. On detecting a reflected photon, the spin is
projected to $|\downarrow \rangle$. To convert the spin from $|\downarrow \rangle$ to $|\uparrow \rangle$
or vice versa, a spin rotation of $\pi$ around y axis is required [see Fig. 2(a) for the definition of
x,y,z axes]. This can be done using a ps or fs optical $(\pi)_y$ pulse (injected from the side of
cavity \cite{muller07, ates09}) via the optical Stark effect. \cite{gupta01, berezovsky08, press08, press10}
To prepare the superposition state such as $|\pm\rangle=(|\uparrow\rangle\pm|\downarrow\rangle)/\sqrt{2}$,
an optical $(\frac{\pi}{2})_y$ pulse can be applied. With these techniques, the electron spin in a
pillar microcavity can be prepared to arbitrary states deterministically.

Spin coherence time $T_2$ is an important parameter for quantum gate and SPT operations in this work.
In GaAs-based or InAs-based QDs, the electron spin dephasing time $T_2^*$ can be quite short ($\sim $ns)
due to the hyperfine interaction between the electron spin and $10^4$ to $10^5$ host nuclear spins. \cite{urbaszek13}
To suppress the nuclear spin fluctuations, spin echo \cite{hahn50} or dynamical decoupling techniques \cite{viola98, viola99}
could be applied to recover electron spin coherence using various pulse sequences made of optical
pulses \cite{gupta01, berezovsky08, press08} and/or single photons. \cite{hu11} Based on the spin echo techniques, $T_2=1\mu$s
for an electron spin has been reported recently in a single In(Ga)As QD. \cite{press10}

Besides the robust deterministic photon-spin entangling gate, this spin-cavity unit \cite{hu09} can also work as
a deterministic photon-photon, spin-spin entangling gate and a photon-spin interface or the heralded spin memory,
single-shot quantum non-demolition measurement of single spin or photon,  complete Bell-state generation,
measurement and analysis  as well as deterministic quantum repeaters. \cite{hu09, bonato10, hu11, wang12}

\section{Photonic transistor}
The linear GCB which is spin dependent can be utilized to make a SPT as shown in Fig. 2(a).
A gate photon sets the spin state via projective measurement and controls the propagation
of single photons or classical optical pulses in the photonic channel between the source S
and the drain D.

The spin is initialized to $|+\rangle=(|\uparrow\rangle+|\downarrow\rangle)/\sqrt{2}$
using single-photon based spin projective measurement in combination with a ultrafast
optical pulse injected from the cavity side (see Sec. II). The gate photon is prepared in an
arbitrary state $|\psi^{ph}\rangle=\alpha |R \rangle+\beta |L\rangle$. After the photon has
interacted with the spin, the joint state turns to the superposition of the transmitted and
reflected states, i.e,
\begin{equation}
-(\alpha|R\rangle^t |\uparrow \rangle+\beta|L\rangle^t |\downarrow \rangle)+(\alpha|R\rangle^r |\downarrow \rangle+\beta|L\rangle^r |\uparrow \rangle). \end{equation}
The photon is then measured in the $\{|H\rangle,|V\rangle\}$ basis with $|H\rangle=(|R\rangle+|L\rangle)/\sqrt{2}$
and $|V\rangle=-i(|R\rangle-|L\rangle)/\sqrt{2}$. On detecting the gate photon in the $|H\rangle$ state
in the transmission port (a click on D2) or reflection port (a click on D4), the spin is projected to
$|\psi^{s}\rangle=\alpha|\uparrow \rangle+\beta|\downarrow\rangle$.

\begin{figure}[ht]
\centering
\includegraphics* [bb= 89 206 469 626, clip, width=7.5cm, height=7.5cm]{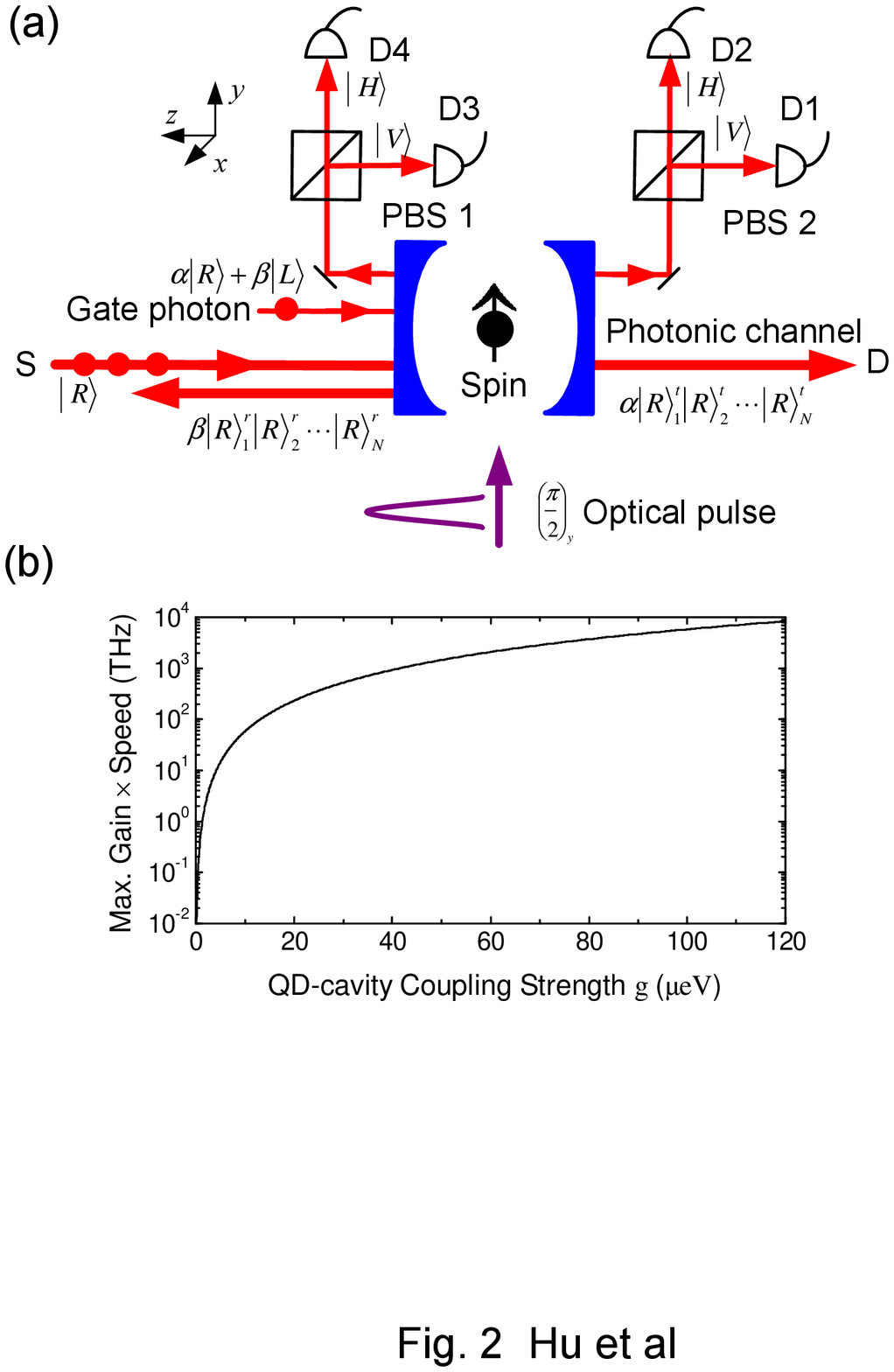}
\caption{(color online). Diagram of the single-photon transistor.
(a) Firstly, the spin is initialized to $|+\rangle=(|\uparrow\rangle+|\downarrow\rangle)/\sqrt{2}$, and the gate
photon is prepared in an arbitrary state $|\psi^{ph}\rangle=\alpha |R \rangle+\beta |L\rangle$;
Secondly, the gate photon state is transferred to spin by measuring the photon in the
$\{|H\rangle,|V\rangle \}$ basis in the transmission and reflection ports; Thirdly,
the spin state is transferred to N single photons (injected in sequence from the source S) by measuring
the spin in the $\{ |+\rangle,|-\rangle \}$
basis. As a result, an arbitrary quantum state of a gate photon is
transferred (or \textquotedblleft amplified\textquotedblright) to the same state encoded on N photons
in the optical channel.
c-PBS (circular polarization beam splitter), D1-D4 (single photon detectors). (b)
The maximum gain - speed product as a function of the QD-cavity coupling strength g. The spin coherence
time is taken as $T_2=1~\mu$s. } \label{fig2}
\end{figure}

Assume there are N photons in the $|R\rangle$ states coming from the source S (the allowed photon number N shall
be discussed later). After all photons have interacted with the spin, the joint state becomes
\begin{equation}
\alpha|R\rangle_1^t|R\rangle_2^t \cdot\cdot\cdot |R\rangle_N^t|\uparrow \rangle+\beta|R\rangle_1^r|R\rangle_2^r \cdot\cdot\cdot |R\rangle_N^r|\downarrow\rangle.
\end{equation}
Note that the $|R\rangle$-components are transmitted, i.e., the photonic channel is open for spin $|\uparrow \rangle$,
or reflected, i.e., the photonic channel is blocked for spin $|\downarrow \rangle$.

An optical $(\frac{\pi}{2})_y$ pulse is then applied from the cavity side to perform the spin Hadamard
transformation (see Sec. II).  After that, a photon in the $|R\rangle$ state is sent to perform spin
measurement, and the N photons are thus projected to a superposition state
\begin{equation}
\alpha|R\rangle_1^t|R\rangle_2^t \cdot\cdot\cdot |R\rangle_N^t \pm \beta|R\rangle_1^r|R\rangle_2^r \cdot\cdot\cdot |R\rangle_N^r,
\label{eq:cat1}
\end{equation}
where \textquotedblleft +\textquotedblright is taken for spin $|\uparrow\rangle$ and
\textquotedblleft -\textquotedblright for spin $|\downarrow\rangle$. The negative sign
can be converted to the positive by guiding one of photons through a $\lambda/2$ waveplate.

As a result, an arbitrary quantum state of a single gate photon is \textquotedblleft amplified\textquotedblright
to the same state encoded on N photons, which is of Greenberger-Horne-Zeilinger (GHZ) state like or
Schr\"{o}dinger-cat state like. \cite{greenberger90} In this sense, this SPT is genuinely
a quantum transistor which shows the dual nature as quantum gate and transistor for entanglement
generation and amplification, respectively. As the original state of gate photon is destroyed after the transistor
operation, this SPT does not violate the no-cloning theorem in quantum mechanics. \cite{wootters82}
This transistor could generate entanglement of hundreds to thousands of photons which
has the potential to surpass the current record of ten-photon entanglement \cite{wang16}.
The multi-photon entanglement generated by this quantum transistor is the key resource for
quantum communications \cite{pan12} and quantum metrology. \cite{giovannetti11}
Previous calculations of the entanglement fidelity in terms of single-photon transportation \cite{hu09} can be extended
to the case of n photons in Fock state ($n\leq P_{max}$) as long as the linear GCB preserves when the incoming power is
less than $P_{max}$.  The fidelity to generate the N-photon GHZ- or cat-like states in Eq.(\ref{eq:cat1})
depends on cavity QED parameters ($g$, $\kappa$, $\kappa_s$, $\gamma$), the spin coherence time $T_2$,
the time interval between photons as well as the spin manipulations and measurement. Detailed discussions
will be presented in future publications.

On detecting the gate photon in the $|V\rangle$ state in the transmission port (a click on D1) or
reflection port (a click on D3), the spin is projected to $|\psi^{s}\rangle=\alpha|\uparrow \rangle-\beta|\downarrow\rangle$.
Following the same procedure as above, the N photons are projected to the same superposition state as
equation (\ref{eq:cat1}) except that the positive sign is for spin $|\downarrow\rangle$ and the
negative sign for spin $|\uparrow\rangle$. This indicates that the SPT works
deterministically.

The time interval between the channel and gate photons should be less than by the spin coherence
time $T_2$ which defines the time window for the transistor operation. Note that the source S and
the drain D are interchangeable as long as the spin quantization direction is changed correspondingly.

The photon rate in the channel can go as high as $P_{max}=\mathrm{g}^2\gamma_{\parallel}/8\kappa \gamma(2\kappa+\kappa_s)$
up to which the linear GCB preserves. \cite{hu15} As the channel opening and closing can be controlled by a single
photon, the maximum gain of the SPT is exactly the maximum photon number allowed in the channel, i.e.,
\begin{equation}
G_{max}=\frac{P_{max}T_2}{\tau},
\end{equation}
where $\tau$ is
the cavity lifetime which determines the cut-off speed. The maximum gain-speed product is thus
\begin{equation}
G_{max}\times \mathrm{Speed}=\frac{\mathrm{g}^2T_2(1+\kappa_s/2\kappa)}{4\hbar^2(1+2\gamma^*/\gamma_{\parallel})},
\end{equation}
where $\hbar$ is the reduced Planck constant. The gain-speed product increases with increasing the coupling
strength g or the spin coherence time $T_2$ as shown in Fig. 2(b) where $T_2=1~\mu$s is taken for an electron
spin in a single In(Ga)As QD. \cite{press10} In the state-of-the-art pillar microcavity \cite{reithmaier04, reitzenstein07}
where $\mathrm{g}/(2\kappa+\kappa_s)=2.4$ ($\mathrm{g}=80~\mu$eV, $2\kappa+\kappa_s=33~\mu$eV),
the maximum gain can reach $7 \times 10^4$, which surpasses other SPT protocols by several orders of
magnitude. \cite{chang07, faraon08, volz12, hwang09, chen13, gorniaczyk14, tiarks14}
A high gain is at the cost of a low speed ($\sim 50~$GHz in this case) and vice versa. In order to raise
the speed, the cavity lifetime can be reduced. For example, if the cavity decay rate increases to $660~\mu$eV,
the speed goes up to  $1~$ THz with the gain down to $3.5\times 10^3$. However, too much cavity decaying
will wash out GCB \cite{hu09} if $4\mathrm{g}^2/(2\kappa+\kappa_s)<\gamma$, and leads to the failure of SPT.

Besides quantum nature, the SPT can also work as a classical optical transistor:  a gate photon in a mixed state
can be amplified to the same mixed state of N photons. Moreover, this classical transistor works either
if a gate photon is replaced by a classical optical pulse as long as the optical power is much less
than $P_{max}$. It is worthy to note that this spin-cavity photonic transistor satisfies all criteria for
optical transistors that compete with the electronic counterparts: \cite{miller10}
cascadability , logic level independent of loss, fan-out, logic-level restoration, absence of critical biasing,
and input/output isolation.

Different from the spin-based electronic transistor based on Rashba spin-orbit interactions, \cite{datta07}
this spin-based photonic transistor based on spin-cavity interactions does not suffer from
the limitation from the RC time constants and the transit time, so it has the potential to break the
THz barrier for all electronic transistors including the state-of-the-art HEMTs. \cite{schwierz07, deal11}

Based on the transistor operation, the spin-cavity unit can be configured to various high-speed (up to THz)
photonic devices, e.g., switches or modulators which are key components for
future Internet technology.

\begin{figure}[ht]
\centering
\includegraphics* [bb= 95 180 485 819, clip, width=6cm, height=9.5cm]{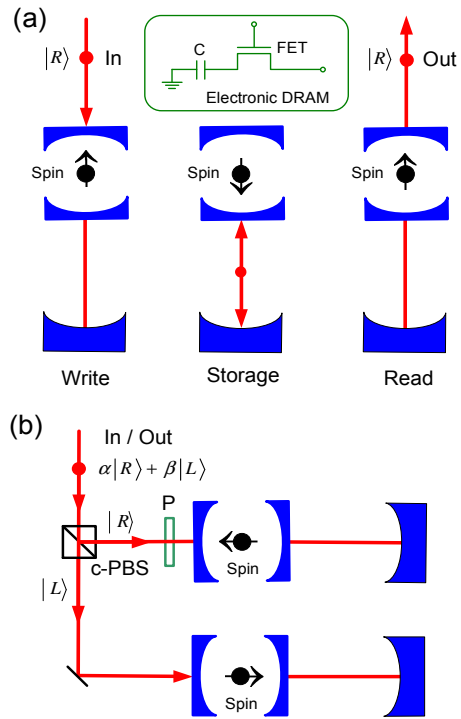}
\caption{(color online). Diagram of the photonic DRAMs. (a) In an optical DRAM, a single photon,
or multiple photons, or classical optical pulses in the $|R\rangle$ or $|L\rangle$ states
can be loaded, stored and unloaded to/from the optical resonator by controlling the spin state.
The inset shows a standard electronic DRAM for comparison. (b) In a quantum photonic DRAM, an
arbitrary single-photon state can be stored and read out to/from two optical DRAMs.
c-PBS (circular polarization beam splitter), P (phase shifter or delay line).  }\label{fig3}
\end{figure}

\section{Photonic DRAM}
In analogue to an electronic capacitor that stores charge, an optical resonator with two highly reflective
end mirrors can store  photons. If one end mirror is replaced with a spin-cavity unit, the spin can be
used as a valve to control the loading, storing, and reading out of photons to/from the optical
resonator [see Fig. 3(a)]. This device is an optical DRAM.

If a photon in the $|R\rangle$ state hits the spin-cavity unit with the spin prepared in the $|\uparrow\rangle$ state,
the spin-cavity unit is transparent, and the photon passes through the spin-cavity unit into the optical resonator.
Immediately, the spin is flipped to $|\downarrow\rangle$ using a ultra-fast optical $(\pi)_y$ pulse (see Sec. II),
and the spin-cavity unit turns to a highly reflective mirror. As a result, the photon resides in the optical resonator.
For reading out, the spin is flipped back to $|\uparrow\rangle$ by applying another ultrafast optical $(\pi)_y$
pulse and the spin-cavity unit becomes transparent again, so the photon passes through the spin-cavity unit
and goes out of the optical resonator.

Depending on the length of resonator and losses on two end mirrors, the photon storage time could reach the
nanosecond or microsecond range. Besides single photon, this optical DRAM can also store multiple photons
as long as the input photon rate is less than $P_{max}$. This optical DRAM is exactly a long-sought
device - optical buffers for all-optical packet switches and all-optical networks. \cite{saleh12}

Fig. 3(b) shows a diagram of a quantum photonic DRAM which can store a quantum state of a single photon.
It consists of two optical DRAMs combined with a c-PBS. The input photon
state $|\psi^{ph}\rangle=\alpha|R\rangle+\beta |L\rangle$ is split into two parts by the c-PBS: $\alpha|R\rangle$
and $\beta |L\rangle$, which are stored in two DRAMs. To read out the state, the two spins are reversed
simultaneously, so the photon comes out and the stored states are combined to the original state via
the c-PBS. The time difference between two paths can be erased by a phase shifter or delay line.

During the whole write-store-read cycle, the spin is always in a basis state $|\uparrow\rangle$
or $|\downarrow\rangle$.  The spin decoherence or relaxation could be overcome by the quantum Zeno
effect \cite{misra77, itano90} if the spin is measured continuously with single photons.
Alternatively, spin echo techniques could be used (see Sec. II).

\begin{figure}[ht]
\centering
\includegraphics* [bb= 56 462 496 681, clip, width=7cm, height=3.5cm]{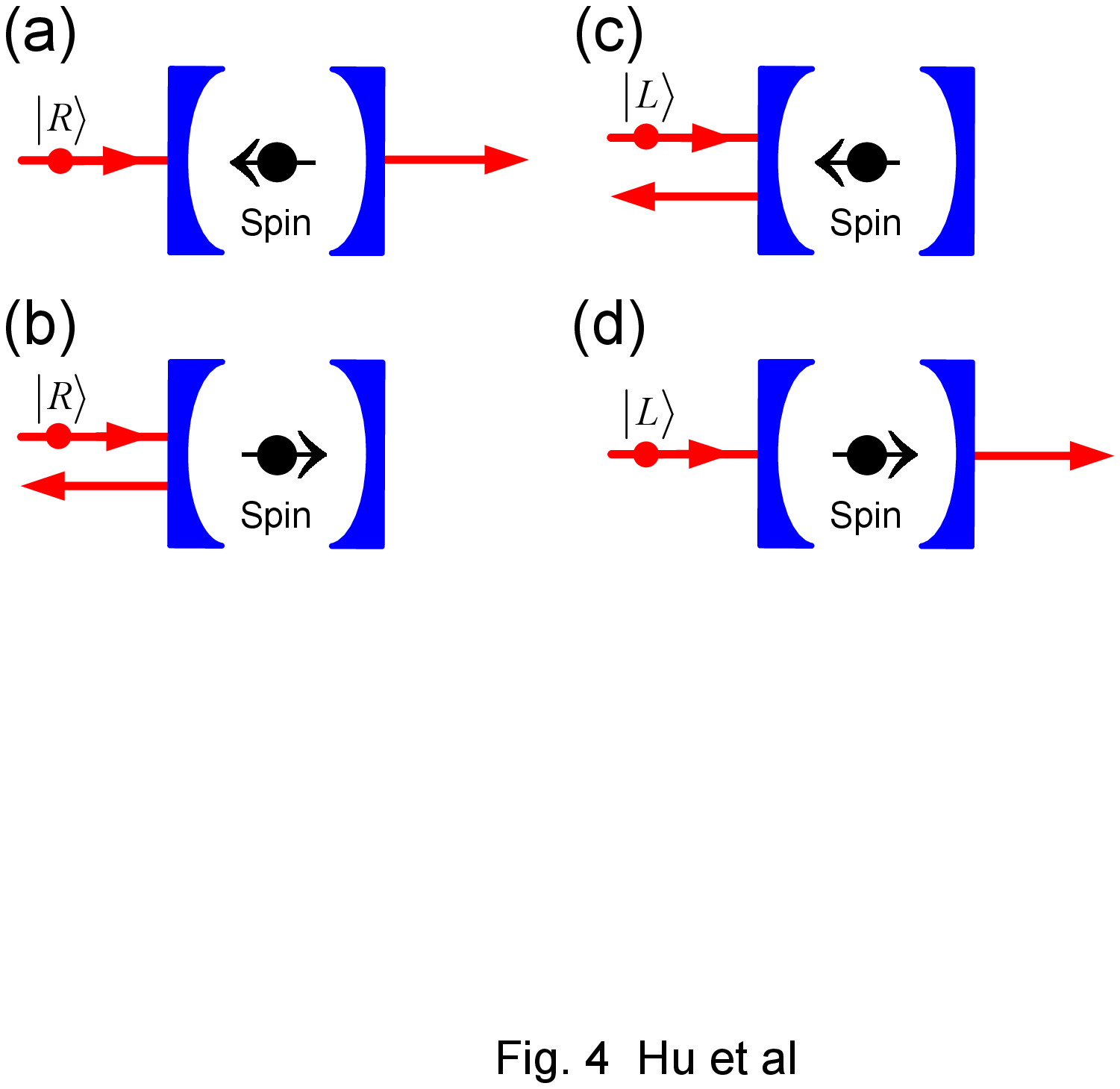}
\caption{(color online). Diagram of the photonic diode. (a) ON; (b) OFF; (c) OFF; (d) ON.} \label{fig4}
\end{figure}

\section{Photonic diode}
The inclusion of a spin into the cavity breaks the time inversion symmetry
of the system if the spin orientation is fixed. This leads to the optical
non-reciprocity, \cite{landau60, agranovich84, potton04} which
can be exploited to make the optical diode or isolator. It is worthy to note here that
GCB is actually a kind of magnetic optical gyrotropy (or magnetic optical activity) which is non-reciprocal,
in contrast to the natural optical gyrotropy (or natural optical activity) in chiral molecules or chiral structures which
is reciprocal. Moreover, optical non-reciprocity is not equivalent to unidirectional light propagation  which
is caused by the breaking of spatial inversion symmetry.
Optical non-reciprocity leads to unidirectional light propagation, but unidirectionality
does not necessarily induce non-reciprocity, \cite{lira12, yin13} e.g., in parity-time synthetic
materials. \cite{bender98, feng13}

If the spin is set to $|\uparrow \rangle$ (pointing from right to left),
a $|R\rangle$-photon from the left will be transmitted to the right, so the diode is on [see Fig. 4(a)].
A $|L\rangle$-photon from the left will be reflected  back to the left, and the diode is off [see Fig. 4(c)].

If the spin is set to $|\downarrow \rangle$ (pointing from left to right),
a $|R\rangle$-photon from the left will be reflected back to the left, so the diode is off [see Fig. 4(b)].
A $|L\rangle$-photon from the left will be transmitted to the right, and the diode is on [see Fig. 4(d)].

As the linear GCB is robust against input power variations, the optical diode can also work at higher powers
up to $P_{max}$. Besides optical diode, the spin-cavity unit can be also configured to other nonreciprocal
devices such as optical isolators or optical circulators \cite{landau60, agranovich84, potton04} for applications
in optical and quantum networks.

\section{Photonic router}
In analogue to classical routers in classical networks and regular Internet, which direct the
data signal to its intended destination according to control information contained in IP address,
quantum routers \cite{lemr13} are a key building block in quantum networks and quantum Internet,
which direct a signal quantum bit (qubit) to its desired output port controlled by the state
of a control qubit, but the signal qubit state is unchanged.

The spin-cavity unit is an ideal component to make a quantum router (see Fig. 5).
The photon is used as the signal qubit in a state $|\psi_s\rangle=\alpha|R\rangle+\beta|L\rangle$
to be directed to its destination, and the spin is used as the control qubit
in state $|\psi_c\rangle$.

If the control spin is set to $|\psi_c\rangle=|\uparrow\rangle$, the $|R\rangle$-component of the signal
photon is transmitted and the $|L\rangle$-component is reflected [see Fig. 5(a)]. The reflected and
transmitted components are then combined by a c-PBS into port c. A phase shifter or delay line  is used
to erase the time different between the $|R\rangle$- and  $|L\rangle$-components such that the output
state in port c remains the same as the original signal state.

\begin{figure}[ht]
\centering
\includegraphics* [bb= 116 207 470 767, clip, width=6cm, height=11cm]{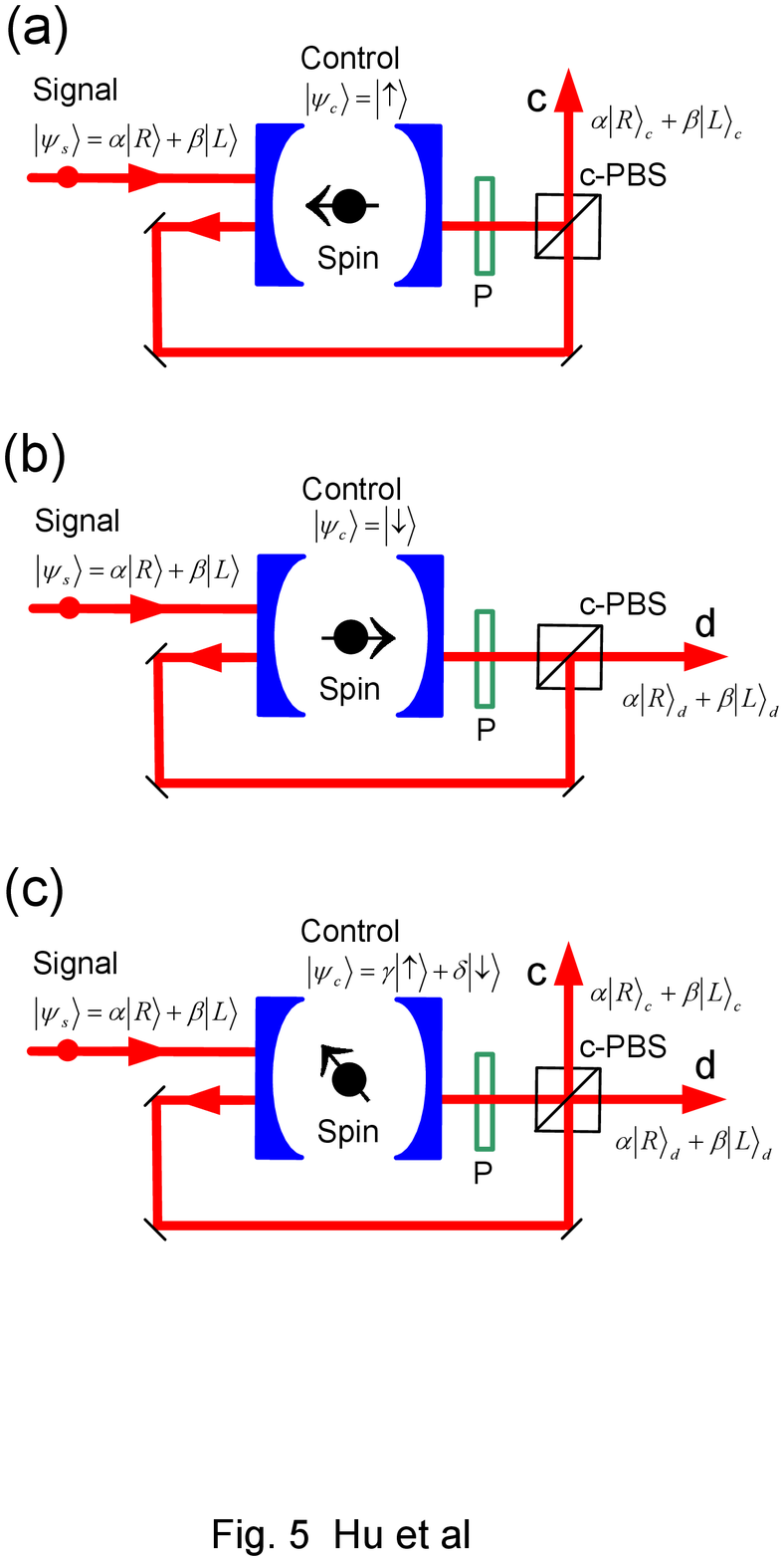}
\caption{(color online). Diagram of the photonic router. The control spin directs the signal photon to: (a) port c;
(b) port d; (c) superposition of two modes in port c and d.
c-PBS (circular polarization beam splitter), P (phase shifter or delay line). } \label{fig5}
\end{figure}

If the control spin is set to $|\psi_c\rangle=|\downarrow\rangle$, the $|L\rangle$-component of the
signal photon is transmitted and the $|R\rangle$-component is reflected [see Fig. 5(b)]. The reflected
and transmitted components are then combined by a c-PBS into port d. The signal photon state remains
unchanged.

If the control spin is set to $|\psi_c\rangle=\gamma|\uparrow\rangle+\delta|\downarrow\rangle$,
the transmitted state is a photon-spin entangled state $\alpha \gamma |R\rangle|\uparrow\rangle+\beta\delta|L\rangle|\downarrow\rangle$, and the
reflected state is another entangled state $\alpha\delta|R\rangle|\downarrow\rangle+\beta\gamma|L\rangle|\uparrow\rangle$ [see Fig. 5(c)].
After combination at the c-PBS, the output state becomes
\begin{equation}
\gamma|\uparrow\rangle(\alpha|R\rangle_c+\beta|L\rangle_c)+\delta|\downarrow\rangle(\alpha|R\rangle_d+\beta|L\rangle_d)
\end{equation}
which is generally a superposition state of two modes in port c and d. The signal photon state
is still unchanged, but can be directed to port c, port d or both controlled by the spin state.
As the spin-cavity unit also works as a photon-spin interface, \cite{hu09} the control spin can
be replaced by a photon, and the quantum router becomes fully transparent. Compared with the
probabilistic quantum router based on linear optics, \cite{lemr13} this spin-cavity quantum router
is deterministic and scalable to multiple photons.

Besides quantum router, the spin-cavity unit can also work as a classical router if single photons
are replaced by classical optical pulses as long as the light power is below $P_{max}$.
This further proves the duality of the spin-cavity unit as quantum gate and transistor.

It is worth pointing out that the spin-cavity unit can also be used to route light (or photons) carrying orbital angular
momentum (OAM) \cite{chen16} if OAM is converted to circular polarization of
light or photons via a q-plate \cite{marrucci06}.

\section{Conclusions and outlook}

The spin-based quantum transistor and related devices discussed above can be made in parallel based
on giant Faraday rotation in another type of spin-cavity unit with a single-sided microcavity. \cite{hu08a, hu08b, hu17}
A unusual feature of these spin-cavity units is the duality as quantum gates and transistors.
On the one hand, the spin-cavity units can work as quantum processors, quantum memories or DRAMs,
quantum repeaters and quantum routers, all of which are key quantum technology for quantum computers
and quantum networks. On the other hand, the spin-cavity units can be configured as optical transistors
for optical information processing and optical buffering with high speed (up to THz). This work demonstrates
that the spin-cavity units provide a solid-state platform ideal for future green and secure Internet - a
combination of all-optical Internet \cite{saleh12} with quantum Internet, \cite{kimble08} which is very
likely to happen within the next 10-20 year timescale. This work series opens up a
new research area - spin photonics.

\appendix

\section{Semiclassical model}
The Heisenberg equations of motions \cite{walls94} for the cavity field operator $\hat{a}$ and the QD dipole
operators $\sigma_-$, $\sigma_z$, together with the input-output relation \cite{gardiner85} can be written as
\begin{equation}
\begin{cases}
& \frac{d\hat{a}}{dt}=-\left[i(\omega_c-\omega)+\kappa+\frac{\kappa_s}{2}\right]\hat{a}-\text{g}\sigma_- \\
& ~~~~~~ -\sqrt{\kappa}\hat{a}_{in}-\sqrt{\kappa}\hat{a}^{\prime}_{in}\\
& \frac{d\sigma_-}{dt}=-\left[i(\omega_{X^-}-\omega)+\frac{\gamma}{2}\right]\sigma_--\text{g}\sigma_z\hat{a} \\
& \frac{d\sigma_z}{dt}=2\text{g}(\sigma_+\hat{a}+\hat{a}^{\dag}\sigma_-)-\gamma_{\parallel}(1+\sigma_z) \\
& \hat{a}_{out}=\hat{a}_{in}+\sqrt{\kappa}\hat{a} \\
& \hat{a}^{\prime}_{out}=\hat{a}^{\prime}_{in}+\sqrt{\kappa}\hat{a},
\end{cases}
\label{eqd1}
\end{equation}
where all the parameters here have the same definitions and meanings as in Eq. (\ref{eqd0}).

If the correlations between the cavity field and the QD dipole are neglected (this is called the semiclassical
approximation), \cite{allen87, armen06} $\langle \sigma_{\pm}\hat{a}\rangle =\langle \sigma_{\pm}\rangle\langle \hat{a} \rangle$
and $\langle \sigma_z\hat{a}\rangle =\langle \sigma_z\rangle\langle \hat{a} \rangle$. The
semiclassical approximation can be applied in three cases: \cite{hu15}
(1) low-power limit $P_{in}\ll 1$ where the QD is in the ground state (weak-excitation approximation);
(2) high-power limit $P_{in}\gg 1$ where the QD is strongly saturated; (3) within the non-saturation
window where the QD stays in the ground state. The reflection and transmission coefficients can thus be
derived as
\begin{equation}
\begin{split}
& r(\omega)=1+t(\omega), \\
& t(\omega)=\frac{-\kappa[i(\omega_{X^-}-\omega)+\frac{\gamma}{2}]}{[i(\omega_{X^-}-\omega)+
\frac{\gamma}{2}][i(\omega_c-\omega)+\kappa+\frac{\kappa_s}{2}]-\text{g}^2 \langle \sigma_z\rangle}.
\end{split}
\label{eqd2}
\end{equation}

The population difference  $\langle\sigma_z\rangle$ is given by
\begin{equation}
\langle\sigma_z\rangle=-\frac{1}{1+\frac{\langle n\rangle}{n_c[1+4(\omega_{X^-}-\omega)^2/\gamma^2]}},
\label{eqd3a}
\end{equation}
and the average cavity photon number $\langle n\rangle \equiv \langle \hat{a}^{\dagger}\hat{a}\rangle$ by
\begin{widetext}
\begin{equation}
\langle n\rangle=\frac{\kappa[(\omega_{X^-}-\omega)^2+\frac{\gamma^2}{4}]P_{in}}
{[(\omega_{X^-}-\omega)^2+\frac{\gamma^2}{4}][(\omega_c-\omega)^2+\frac{(2\kappa+\kappa_s)^2}{4}]+2\text{g}^2\langle\sigma_z\rangle
[(\omega_{X^-}-\omega)(\omega_c-\omega)-\frac{(2\kappa+\kappa_s)\gamma}{4}]+\text{g}^4\langle\sigma_z\rangle^2},
\label{eqd3b}
\end{equation}
\end{widetext}
where $n_c=\gamma_{\parallel}\gamma/8\mathrm{g}^2$ is the critical photon number which measures the average
cavity photon number required to saturate the QD response.
$P_{in}=\langle\hat{a}^{\dagger}_{in}\hat{a}_{in}\rangle$ is the incoming light power.
$\langle\sigma_z\rangle$ is the QD population difference between the excited state and the
ground state, and can be used to measure the saturation degree. $\langle\sigma_z\rangle$
ranges from $-1$ to $0$. If $\langle\sigma_z\rangle=-1$, the QD is in the ground state (not saturated);
if $\langle\sigma_z\rangle=0$, QD is fully saturated, i.e., $50\%$ probability in the ground state and $50\%$
probability in the excited state. If $\langle\sigma_z\rangle$ takes other values, the QD is partially saturated.

By solving Eqs. (\ref{eqd3a}) and (\ref{eqd3b}),  $\langle\sigma_z\rangle$ and $\langle n\rangle$
can be obtained at any input power. Note that $\langle\sigma_z\rangle$ and $\langle n\rangle$ are
dependent on the input power, frequency $\omega$ and coupling strength g. Putting $\langle\sigma_z\rangle$
into Eq. (\ref{eqd2}), the reflection and transmission coefficients can be obtained.

The linear GCB preserves as long as the the non-saturation window between the resonances of two first-order
dressed states (or two polariton states) is open.\cite{hu15}  From Eqs. (\ref{eqd3a}) and  (\ref{eqd3b})
it can be derived that the non-saturation window is closed roughly at
$P_{max}=\mathrm{g}^2\gamma_{\parallel}/8\kappa \gamma(2\kappa+\kappa_s)$ (normalized
by photons per cavity lifetime) where $\langle\sigma_z\rangle=-1/2$ is reached.
As the non-saturation window is a highly reflective region,  the higher is the coupling
strength g, the higher powers the linear GCB can preserve.

\section{Full quantum model - master equation}
The reflection and transmission coefficients can be also calculated numerically in the frame of master
equations in the Lindblad form \cite{walls94} by using a quantum optics toolbox. \cite{tan99, johansson13}
The master equation for the spin-cavity system can be written as
\begin{equation}
\begin{split}
\frac{d\rho}{dt}= &-i[H_{JC},\rho]+(\kappa+\kappa_s)(\hat{a}\rho \hat{a}^{\dag}-\frac{1}{2}\hat{a}^{\dag}\hat{a}\rho-\frac{1}{2}\rho\hat{a}^{\dag}\hat{a})\\
& +\gamma_{\parallel}(\hat{\sigma}_-\rho \hat{\sigma}_+ - \frac{1}{2}\hat{\sigma}_+\hat{\sigma}_-\rho-\frac{1}{2}\rho\hat{\sigma}_+\hat{\sigma}_-)+
\frac{\gamma^*}{2}(\hat{\sigma}_z\rho\hat{\sigma}_z-\rho)\\
\equiv & \mathcal{L}\rho,
\end{split}\label{master1}
\end{equation}
where the parameters $\kappa, \kappa_s, \gamma, \gamma_{\parallel}, \gamma^{*}$ are defined in the same way as
in Eq. (\ref{eqd1}), $\mathcal{L}$ is the Liouvillian and $H_{JC}$ is the driven Jaynes - Cummings Hamiltonian
with the input field driving the cavity. In the rotating frame at the frequency of the input field, $H_{JC}$
can be written as
\begin{equation}
\begin{split}
H_{JC}= &(\omega_c-\omega)\hat{a}^{\dagger}\hat{a}+(\omega_{X^-}-\omega)\sigma_+\sigma_-\\
& +ig(\sigma_+\hat{a}-\hat{a}^{\dagger}\sigma_-)+i\sqrt{\kappa}\hat{a}_{in}(\hat{a}-\hat{a}^{\dagger}),
\end{split}
\end{equation}
where  the input field is associated with the output field and the cavity field by the input-output
relation, \cite{gardiner85} $\hat{a}_{out}=\hat{a}_{in}+\sqrt{\kappa}\hat{a}$.

Although an analytical solution to the master equation in Eq. (\ref{master1}) is very difficult,
A quantum optics toolbox in Matlab \cite{tan99} or in Python \cite{johansson13} provides an exact
numerical calculation of the density
matrix $\rho(t)$. By taking the operator average in the input-output relation, the reflection
and transmission coefficients in the steady state can be calculated by the following expression
\begin{equation}
\begin{split}
& r(\omega)=1+t(\omega), \\
& t(\omega)=\sqrt{\kappa}\frac{\mathrm{Tr}(\rho \hat{a})}{\langle\hat{a}_{in}\rangle}.
\end{split}
\end{equation}

\section*{ACKNOWLEDGMENTS}
The author thanks the financial support of EPSRC fellowship EP/M024458/1.

\end{document}